\documentclass[12pt,preprintnumbers]{article}
\usepackage[utf8]{inputenc}
\usepackage{amsmath, amssymb, wasysym, slashed, multirow,hyperref}
\usepackage{pdfpages}
\usepackage{cite}
\usepackage{times}

\newenvironment{sciabstract}{%
\begin{quote} }
{\end{quote}}

\newcounter{lastnote}

\usepackage{fancyhdr}
\fancypagestyle{plain}{%
	\fancyhead[R]{ACFI-T21-12}

}

\title{The QCD Adler function and the muon $g-2$ anomaly from renormalons}

\author
{Alessio Maiezza,$^{1,2\ast}$  Juan Carlos Vasquez$^{3,4\dagger}$\\
\\
\normalsize{$^{1}$Dipartimento di Scienze Fisiche e Chimiche, Universit\`a} \\
\normalsize{degli Studi dell'Aquila, via Vetoio, I-67100, L'Aquila, Italy,}\\ \\
\normalsize{$^{2}$Ruder Bo\v skovi\'c Institute, Bijeni\v cka cesta 54, 10000, Zagreb, Croatia.}\\ \\
\normalsize{$^{3}$Department of Physics and Astronomy, Science Center, }\\ 
\normalsize{Amherst College, Amherst, MA 01002, USA.}\\ \\
\normalsize{$^{4}$Amherst Center for Fundamental Interactions, Department of Physics,}\\ 
\normalsize{University of Massachusetts, Amherst, MA 01003, USA.}\\
\\
\small{ E-mail: alessiomaiezza@gmail.com$^{\ast}$,jvasquezcarm@umass.edu$^{\dagger}$}
}

\date{}

\begin{document}


\baselineskip16pt %24 is the original


\maketitle


\begin{sciabstract}
We describe the Adler function in Quantum Chromodynamics using a transseries representation within a resurgent framework. The approach is based on a Borel-Ecalle resummation of the infrared renormalons combined with an effective running for the strong coupling. The new approach is flexible enough to give values in agreement with the current Adler function determinations. We then apply our finding to the muon's anomalous magnetic moment studying the possibility of saturating, solely in terms of the vacuum polarization function, the current discrepancy between the best Standard Model value for the muon's anomalous magnetic moment and the experimental value obtained by the most recent muon $g-2$ collaboration. The latter shows that the Adler function's new representation can also be consistent with recent lattice determinations.
\end{sciabstract}

% -1 General Introduction: Adler function, non-perturbative methods
\section{Introduction}

The Quantum Chromodynamics (QCD) description at the hadronic scale is a formidable challenge due to the breakdown of the perturbation theory description for finite values of the coupling constant. Currently, perturbation theory is the  \emph{only analytical} tool to compute physical quantities within quantum field theory (QFT). The Adler function~\cite{PhysRevD.10.3714} is a fundamental quantity used in QCD to describe the nonperturbative effects at the hadronic scale. Its perturbative expression is known for up to five loops~\cite{Baikov:2008jh}. Its theoretical description is essential since it appears in any process involving QCD corrections due to the vacuum hadronic polarization function. Lattice QCD allows performing a nonperturbative treatment of the QCD Adler function at the hadronic scale~\cite{Francis_2013}. However, analytic, nonperturbative solutions are hard or impossible to obtain within the lattice framework. Hence, an analytical understanding would be beneficial. Based on the notion of renormalons~\cite{PhysRevD.10.3235,LAUTRUP1977109,tHooft:1977xjm,Parisi:1978iq} and Operator-Product-Expansion~\cite{wilson1972,Shifman:2021iis}, there are non-perturbative analytical evaluations for the Adler function and QCD observables~\cite{Beneke:1998ui,Shifman:2013uka,Cvetic:2018qxs,Caprini:2020lff}. Other methods use integral representations~\cite{Shirkov:1997wi,Nesterenko:2007fm,Cvetic:2008bn}. Although all these analytical approaches reproduce some qualitative features of the ``experimental'' Adler function~\cite{Peris:1998nj}, the description is insufficient to describe the experimental data at the hadronic scale.

% -2 Us within resurgent, and introduction on the resurgence in general (Ecalle, etc..)
The recent analytical approach to the Adler function of Ref.~\cite{Maiezza:2021mry} distinguishes itself from previous ones because it is based on renormalons and the resurgence theory. First proposed by Ecalle~\cite{Ecalle1993} in a purely mathematical context, it has found fertile ground in QFT~\cite{Argyres:2012ka,Dunne:2012ae,Dorigoni:2014hea,Aniceto:2018bis,Clavier:2019sph,Borinsky:2020vae,Fujimori:2021oqg,Costin:2019xql,Costin:2020hwg,Borinsky:2022knn}. Renormalized perturbation theory controls the finiteness of QFT in the proper regime. Therefore, it is an appealing possibility to continue it to the nonperturbative regime analytically. For this reason, resurgence may represent a good candidate for a foundational, analytical approach to a nonperturbative QFT.

% -3 From general resurgence to specific resurgence (Costin). An introduction to free parameters (important for later!)
In the specific framework of ordinary-differential-equations~\cite{Costin1995,CostinBook} (ODEs), a resurgent approach to the renormalization group was proposed in Refs~\cite{Maiezza:2019dht,Bersini:2019axn}. The renormalization group equation (RGE) in this new approach is written as a non-linear ODE in the coupling constant. The resulting theory is then applied to the QCD Adler function~\cite{Maiezza:2021mry}, where the renormalons can be resummed, leaving one arbitrary constant fixed from data. It represents an improvement to all the known renormalon-based evaluations in QFT and QCD. The inability to calculate this arbitrary constant is due to the non-existence of a semiclassical limit for renormalons. Because of the technical details we shall discuss, the transseries representation for the Adler function has three arbitrary constants to be determined from the data.

% -4 Last, the result (for Adler only)!
In this work, we show that the new approach of Ref.~\cite{Maiezza:2021mry} has the flexibility to reproduce the Adler function data in the entire infrared (IR) regime, provided that one properly regularizes the Landau pole~\cite{Landau} singularity. To this aim, we adopt an effective running for the strong coupling $\alpha_s$ that prevents the coupling from diverging~\cite{Cornwall:1981zr}. The nonperturbative running we adopt is such that the strong coupling freezes at low energy~\cite{PhysRevD.44.1285} -- see the review~\cite{Deur:2016tte} for typical nonperturbative running for $\alpha_s$. The final result is shown in Fig.~\ref{Adler}. Our result features three parameters in contrast to conventional renormalon approaches with an infinite number of arbitrary constants. To illustrate the predictivity of the transseries representation, we also compare it with a fit including the same number of free parameters (three) but in conventional renormalon-based evaluation. In this case, there is no agreement between the theory and data for energies below $\approx 1.3$ GeV.

% -5 Finally, application to g-2
We then apply the new tool to the $g-2$ discrepancy~\cite{Miller:2007kk,Miller:2012opa}. In particular, we follow the possibility considered in Ref.~\cite{Keshavarzi:2020bfy}, in which the QCD vacuum polarization function is tentatively modified below $\sim0.7$ GeV because this is not yet excluded at that energy to saturate the $g-2$ discrepancy.
The latter would agree with the most recent experimental result obtained by the $g-2$ collaboration~\cite{PhysRevLett.126.141801} and the most recent lattice computation~\cite{Borsanyi:2020mff}. From the proposed approach's perspective, we study the impact of the transseries representation of the Alder function on the anomalous magnetic moment of the muon.

\section{The Adler function}

 The Adler function $D(Q)$ is defined as
\begin{equation}
D\left(Q\right)=4 \pi^{2} Q^{2} \frac{\mathrm{d} \Pi\left(Q\right)}{\mathrm{d} Q^2}\,,
\end{equation}
where $\Pi(Q)$ is determined via
\begin{align}\label{twopointfunction_currents}
-\mathrm{i} \int d^{4} x \, & \mathrm{e}^{-i q x}\left\langle 0\left|T\left(j_{\mu}(x) j_{\nu}(0)\right)\right| 0\right\rangle \nonumber \\
&=\left(q_{\mu} q_{\nu}-q^{2} g_{\mu \nu}\right) \Pi\left(Q\right)\,,
\end{align}
being $q$ the transferred momentum, $Q^2=-q^2$ and $j_{\mu}= \bar{q}\gamma_{\mu}q$ two massless quark currents. In perturbation theory, the Adler function is given by
\begin{equation}\label{Adlerpert}
D_{pert}\left(Q\right)=1+\frac{\alpha_{s}}{\pi} \sum_{n=0}^{\infty} \alpha_{s}^{n}\left[d_{n}\left(-\beta_{0}\right)^{n}+\delta_{n}\right]\,.
\end{equation}
We use the convention for the beta function $\beta(\alpha_s)=\mu^2\frac{d\alpha_s}{d\mu^2}= \beta_0\alpha_s^2+\beta_1\alpha_s^3+\mathcal{O}(\alpha_s)^4$, $2\pi\,\beta_0=-11+\frac{2}{3}n_f $, where $n_f$ is the number of active flavors, $\alpha_s = g_s^2/4\pi$ and $g_s$ denotes the gauge coupling of the strong interaction gauge group $SU(3)_c$. The Adler function expression in perturbation theory up to $n=2$ can be found in Refs.~\cite{Gorishnii:1990vf,Surguladze:1990tg,Kataev:1995vh,Beneke:1998ui}, from which the coefficients $d_n$ and $\delta_n$ can be extracted, as shown in Ref.~\cite{Beneke:1998ui}. The five-loop coefficient $d_3$ and $\delta_3$ can be extracted from Ref.~\cite{Baikov:2008jh}. Notice that in Eq.~\eqref{Adlerpert} one needs to know the all coefficients $d_n$ and $\delta_n$ up to $n\rightarrow\infty$. As we discussed, it is only possible to compute the first few orders in perturbation theory expansions. Fortunately, there is a well-known procedure in the literature called ``Naive non-abelianization"~\cite{Beneke:1994qe,Beneke:1998ui}, which is used to estimate Eq.~\eqref{Adlerpert}  to all orders in perturbation theory. Essentially, within this procedure, one estimates the large order behavior in Eq.~\eqref{Adlerpert} using the first know coefficients $d_n$ and $\delta_n$ from perturbation theory. The remaining coefficients are then estimated using the property that $d_n\propto n!$ whereas $\delta_n$ is not, so that for sufficiently large $n$ one has $\delta_n/d_n\rightarrow 0$. As a crude approximation, one then sets the coefficients $\delta_n$  to zero for $n\geq 4$. This procedure is expected to give more accurate results as new perturbative computations are included. Finally, the coefficients $d_n$ for $n\geq 4$ are estimated  from the fermion renormalons graphs computed in Ref.~\cite{Beneke:1998ui}.

\section{Resurgent Adler function} \label{RRGEA}

In Ref.~\cite{Maiezza:2021mry}, we resummed the IR renormalon contribution to the QCD Adler function using the Borel-Ecalle resummation of Refs.~\cite{Bersini:2019axn,Maiezza:2019dht} -- see the appendices~\ref{app:Synthesis}, \ref{ODE_RES} and~\ref{Detail_RRGE_Adler} for more details on the resurgent approach.
After resuming the renormalons using the new framework, the expression for the Adler function features three arbitrary constants: one constant $c_1$ parametrizing simple pole ambiguity due to the first non-zero renormalon; another constant $C$ stemming from the Borel-Ecalle resummation of quadratic renormalons; a constant $K$ related to the $n!$ behavior in the perturbative series~\cite{Beneke:1998ui}, which in the case of renormalons and unlike instantons~\cite{Lipatov:1976rb}, cannot be determined using semiclassical methods~\cite{tHooft:1977xjm}. The inability to determine those constants from first principles is a well-known problem. It has been recently linked to foundational issues to construct an unambiguous QFT starting from the free fields~\cite{Maiezza:2020qib}.

The original fermion bubble graph contribution to the Adler function was calculated in Ref.~\cite{Neubert:1994vb}. In Ref.~\cite{Maiezza:2021mry}, we rewrote the fermion bubble graph contribution of Ref.~\cite{Neubert:1994vb} such that the pole structure of the Borel transform was apparent. After applying the Borel-Ecalle resummation of Refs.~\cite{Bersini:2019axn,Maiezza:2019dht}, the transseries expression of the Adler function is of the form.
\begin{align} \label{trans_adler}
D_{resurg.}(Q) =\, & D_0(Q)-\frac{4\pi}{\beta_0} c_1e^{\frac{2}{\beta_0\,\alpha_s(Q^2)}}\nonumber \\
& + C e^{\frac{1}{\beta_0\,\alpha_s(Q^2)}}\left(\frac{1}{\alpha_s(Q^2)}\right)^{a_p}D_1(Q^2)\,,
\end{align}
where $a_p =1+ \mathcal{O}(\beta_1/\beta_0^2)$. The function $D_0(Q)$ contains the perturbative expression up to $\mathcal{O}(\alpha_s^4)$ shown in Eq.~\eqref{Adlerpert} and is given by:
\begin{equation}\label{D0}
D_0(Q) = D_{pert}(Q)+ D_{K}(Q)\,,
\end{equation}
where $D_{K}(Q)\propto \sum_{n=0}^{\infty}K\,\beta_0^n\, \alpha_s^{n+1}n!$. Following the formalism of Ref.~\cite{CostinBook}, we then regularize the $n!$ divergence in $D_K$ by taking the Cauchy principal value for the Laplace integral such that
\begin{align}\label{Adler_K}
\frac{D_{K}(Q)}{2K}& =\frac{e^{\frac{2}{\alpha_s\beta_0}}
\Gamma \left(0,\frac{2}{\alpha_s\beta_0}\right)}{\beta_0} + \frac{2 e^{\frac{3}{\alpha_s\beta_0}} \Gamma \left(0,\frac{3}{\alpha_s\beta_0}\right)}{3 \alpha_s\beta_0^2}+\nonumber \\
&\sum_{p=1}^{\infty}
 \left(\frac{ \alpha_s\beta_0-2 (p+1)
 e^{\frac{2 (p+1)}{\alpha_s\beta_0}} \Gamma
 \left(0,\frac{2 (p+1)}{\alpha_s\beta_0}\right)}{3\beta_0^2\alpha_sp (p+1) (2 p+1)}+\right. \nonumber\\
 &\left. \frac{2 \left((2 p+3)
 e^{\frac{2 p+3}{\alpha_s\beta_0}} \Gamma
 \left(0,\frac{2 p+3}{\alpha_s\beta_0}\right)-\alpha_s\beta_0\right)}{3\beta_0^2\alpha_s (p+1) (2 p+1) (2 p+3)}\right)\,.
\end{align}
In the above expression, the terms up to $\mathcal{O}(\alpha^4_s)$ must be removed in $D_K(Q)$ to prevent the double counting of this contribution in $D_{pert}(Q)$. The Eq.~\eqref{Adler_K} is derived in the appendix~\ref{Detail_RRGE_Adler}.

Finally, the function $D_1(Q^2)$ is found from $D_0$~\cite{Maiezza:2019dht} using resurgence as shown in Refs.~\cite{Maiezza:2019dht,Bersini:2019axn}. Choosing the renormalization scale $\mu^2= Q^2e^{-5/3}$ and neglecting the two-loop corrections proportional to $\beta_1$, one finds~\footnote{We also proved that Eq.~\eqref{nonpert_adler} could be derived using Ecalle bridge equation obtained from the RGE. We will discuss the latter point in a separate publication.}
\begin{align}
D_1(Q) =& \,\frac{8 \pi K }{3 \alpha_s \beta_0^2}\left[2e^{\frac{1}{\alpha_s\beta_0}} -
 \left(e^{\frac{1}{\alpha_s\beta_0}}+1\right)
 \log \left(1-e^{\frac{2}{\alpha_s \text{$\beta
 $0}}}\right)\right.\nonumber \\ -
 &\left. 2 \left(e^{\frac{1}{\alpha_s \text{$\beta
 $0}}}+1\right) \tanh ^{-1}\left(e^{\frac{1}{\alpha_s
 \beta_0}}\right)\right]\,. \label{nonpert_adler}
\end{align}
The next step is implementing the nonperturbative running for the coupling $\alpha_s(\mu)$ to be used in Eq.~\eqref{trans_adler}.

\section{Effective running and the QCD Adler function at low energies}

In Ref.~\cite{Dokshitzer:1995qm}, the authors explored the possibility that the QCD running coupling can be effectively extrapolated in a process-independent way to smaller momenta of the order of the hadronic scale. The essential idea is that nonperturbative physics should reveal itself smoothly in inclusive observables. Consequently, it is meaningful to extend the notion of the QCD coupling $\alpha_s$ down to zero energy for these types of observables. These arguments apply to the QCD Adler function as well.

The transseries provided in Ref.~\cite{Maiezza:2021mry} can fit the experimental Adler function up to energy $\approx0.7$GeV. The failure below that energy is due to the unphysical Landau pole of the perturbative running of $\alpha_s$ -- and not the Borel-Ecalle resummation formalism. To overcome this difficulty, in this work, we use an effective running coupling valid up to zero energy in which $\alpha_s$ goes to a constant value at zero energy. We should stress here that, in this case, the absence of the IR Landau pole is not in contradiction with the presence of the renormalons and their Borel-Ecalle resummation in Eq.~\eqref{trans_adler}. The correspondence between the Landau pole and renormalons only holds at the perturbative level. In particular, the renormalons are calculated using the one-loop $\beta-$function. The renormalons only signal the nonperturbative energy scale $\Lambda$ at which perturbation theory breaks down, as initially discussed in Ref.~\cite{Parisi:1978bj} and elaborated, among others, in Refs.~\cite{Dokshitzer:1995af,Peris:1996in}. More recently, the issue has been analyzed with the resurgence of the RGE, taking the $\phi^4$-model as a prototype~\cite{Maiezza:2020nbe}, where the possibility that non-analytic corrections from renormalons make the model asymptotically safe is argued~\footnote{The avoidance of Landau pole via non-analytic (flat) contributions was previously discussed in Ref.~\cite{Klaczynski:2013fca}.}.\\

In our specific case, we use Cornwall's coupling~\cite{Cornwall:1981zr}, which is one of the simplest analytic nonperturbative models for the running of $\alpha_s$ and given by~\cite{PhysRevD.44.1285}
\begin{equation}\label{eff_running}
\alpha_s(Q)=\frac{4 \pi}{11 \ln \left(z+\chi_{g}\right)-2 n_{\mathrm{f}} \ln \left(z+\chi_{q}\right) / 3}\,,
\end{equation}
where $z=Q^2/\Lambda^2$, $n_f$ is the number of flavors, $\chi_g = 4m_g^2/\Lambda^2$, $\chi_q=4m_q^2/\Lambda^2$, the light constituent quark mass $m_q= 350$ MeV, the gluon mass $m_g\simeq 500 $ MeV, and $\Lambda$ denotes the QCD hadronic (non-perturbative) scale. We shall determine $\Lambda$ by fitting the experimental data of the Adler function.

A comment is now in order. Our approach provides a representation of the QCD Adler function as a transseries in $\alpha_s$, and the only requirement for Eq.~\eqref{trans_adler} to hold is that the coupling is not too large~\cite{Maiezza:2021mry}. The perturbative and the effective couplings coincide in the UV, where they are sufficiently small for the Eq.~\eqref{trans_adler} to be valid. Therefore, an effective top-bottom running of Eq.~\eqref{trans_adler} is performed through Eq.~\eqref{eff_running}.

\begin{table}
\centering
\tabcolsep=0.05cm
\begin{tabular}{ r | c | c}
\hline
 \hline
  Parameter &  Low energy fit (4-loop) &Low energy fit (5-loop) \\
\hline
\hline
$K$ & 1.422 & 0.805\\
$C$ & 0.629 & 0.240 \\
$c_1$ & 0.0326 & -0.358 \\
 \hline
$\Lambda$& 731 MeV & 697 MeV \\
 \hline \hline
\end{tabular}
\caption{Numerical value of the constants in Eq.~\eqref{trans_adler}. The central column represents the low energy ($Q\lesssim 1.3$ GeV) fit, starting from the four-loop perturbative expression. The third column shows the values starting from the five-loop perturbative expression, corresponding to the plot in Fig.~\ref{Adler}.}
\label{tabrersurgencefit}
\end{table}
Notice also that, based on the (resummation of) renormalons, the result is intrinsically effective because of the inability to determine the parameters $K, C, c_1$. Therefore, the use of Eq.~\eqref{eff_running} for describing the low-energy running of $\alpha_s$ brings in no additional conceptual changes~\footnote{The operation of power corrections to physical observables and make the coupling $\alpha_s$ effective (analyzation) do not commute, and this would lead to ambiguity in Eq.~\eqref{trans_adler}~\cite{Cvetic:2006mk,Cvetic:2006gc,Cvetic:2013gta}. However, since Eq.~\eqref{trans_adler} is intrinsically ambiguous, one can reabsorb the ambiguity mentioned above in the definition of the fitted parameters (e.g., $"C"$).}. We find that the typical running for $\alpha_s$ reproducing the Adler function is such that at low energies, $\alpha_s(0)\simeq 1.6$, which is in the ballpark of known results in the literature. See Ref.~\cite{Deur:2016tte} and references therein for a detailed discussion about the low energy behavior of the QCD running coupling in several nonperturbative approaches. The possibility of describing the nonperturbative running of $\alpha_s$ within the resurgent framework merits dedicated analysis and is therefore left for future work.

In Tab.~\ref{tabrersurgencefit}, we show the values for the parameters entering in the transseries for the Adler function in Eq.~\eqref{trans_adler}. We show the values obtained from the fit using the four-loop and five-loop expressions for the Adler function and find agreement with the current determinations of the QCD Adler function in both cases. However, as seen from the table, there is a significant difference in the numerical values for the constants $K, C, c_1$. These variations estimate the theoretical uncertainties for these constants, which we find to be at least $\sim100\%$. We interpret the large errors as the need to include higher-order perturbative corrections: more perturbative information would correspond to convergent values of the parameters. This is not surprising since the entire resurgent approach developed in Refs.~\cite{Bersini:2019axn,Maiezza:2021mry} starts, by construction, from perturbation theory, which needs to be known at all orders in $\alpha_s^n$. In practice, this is not the case, and it is the theoretical reason behind the large theoretical uncertainties for the transseries parameters reported in Tab.~\ref{tabrersurgencefit}.

\begin{figure}
 \centering
 \includegraphics[width=0.5\columnwidth]{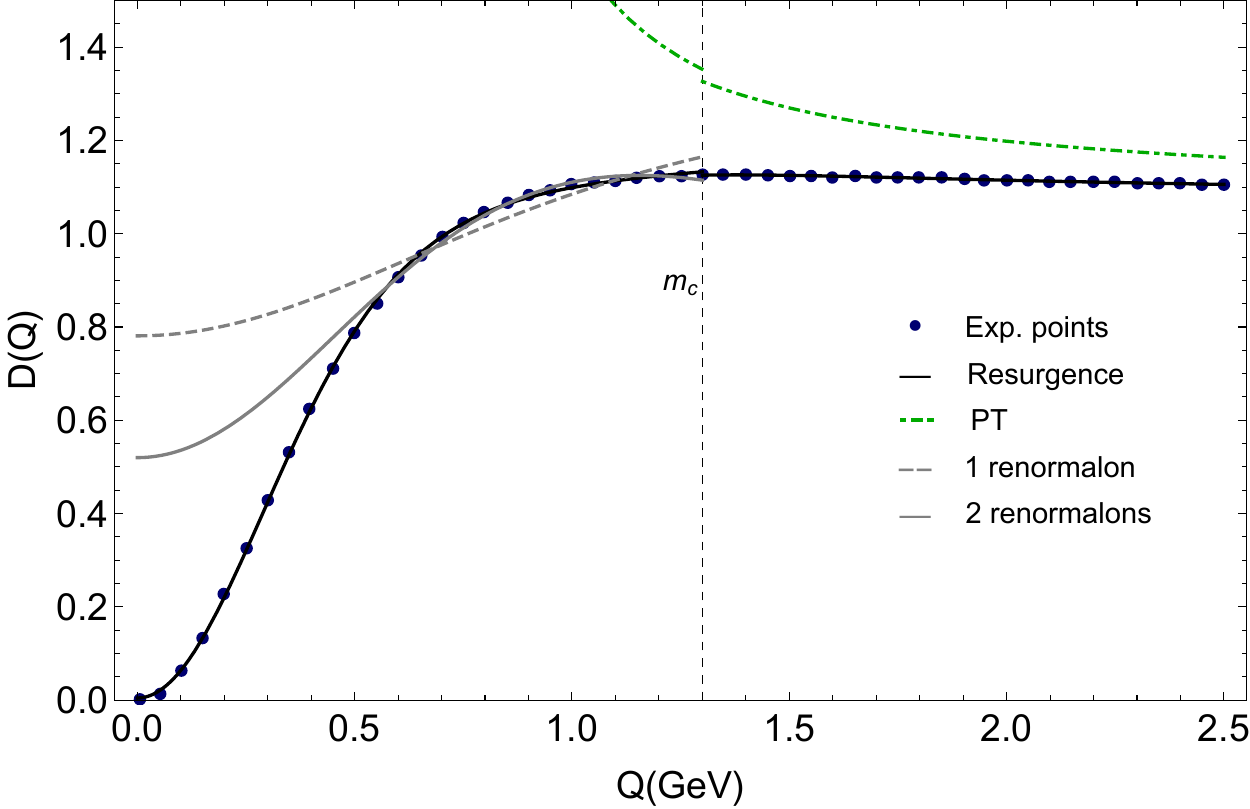}
 \caption{Adler function in the energy range $(0,2.5)$ GeV. The dashed green line is the perturbation theory approximation of the Adler function. Solid Black line corresponds to the resurgent expressions~\eqref{trans_adler} and~\eqref{nonpert_adler}. Dashed and solid gray lines correspond to the approximation of the Adler function, including the first and second renormalon power corrections.}
 \label{Adler}
\end{figure}

With this in mind, we show in Fig.~\ref{Adler} the Adler function in the energy range $Q=(0,2.5)$ GeV using the five loop expression for the Adler function. We see no appreciable difference at energies $Q=(1.3,2.5)$ GeV between the expressions coming from the first two power corrections (gray lines) and the resurgent result (solid black line). Conversely, in the low energy range $Q=(0,1.3)$ GeV, the solid and dashed gray lines fail to describe the Adler function, while the solid black line successfully follows the behavior of the data in the whole range. Despite the significant uncertainties previously discussed, to our knowledge, this is the first time the resurgence formalism provides a phenomenological result for QCD, in particular, an expression for the Adler function that can be used at all energies.

%%%%
%%%%%%%%
%%%%
\section{Saturating the $g-2$ experimental discrepancy of the muon anomalous magnetic moment of the muon} \label{amu}
\begin{figure}[h]
\centering
\includegraphics[width=0.5\columnwidth]{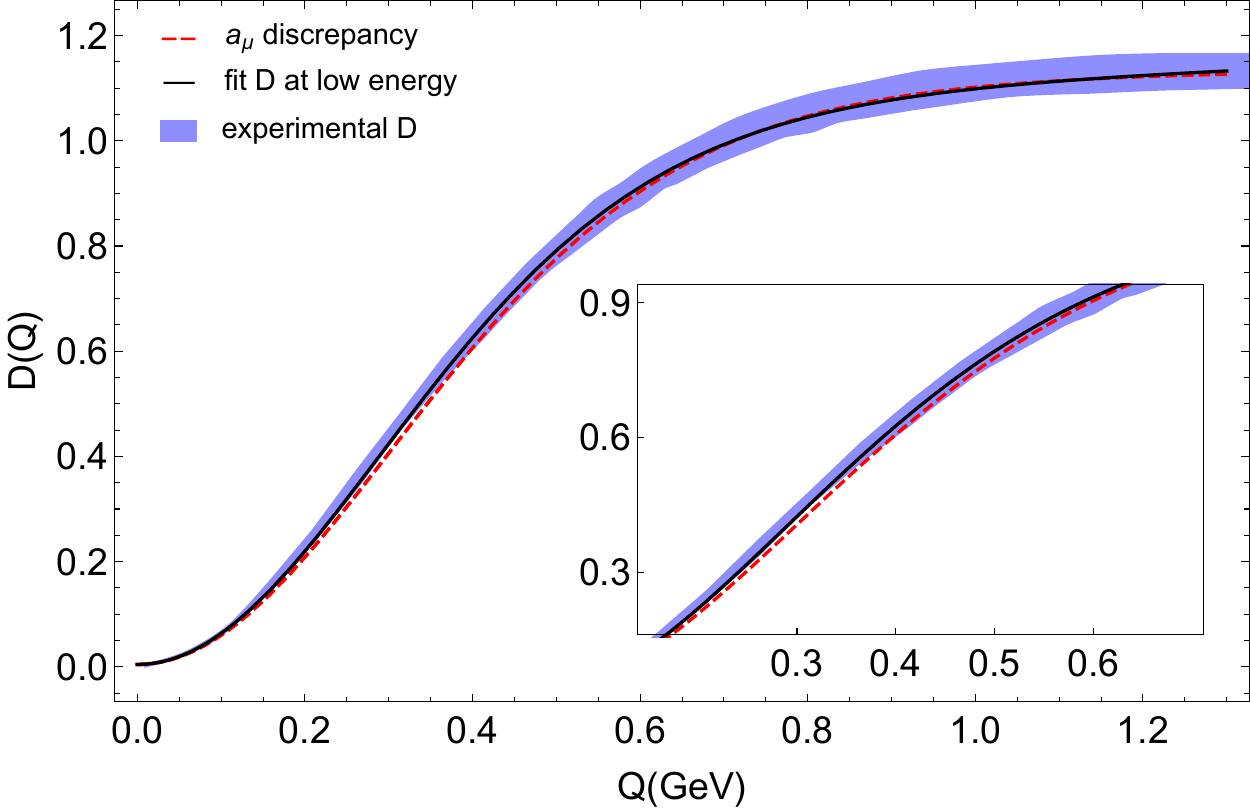}
\caption{The Adler function in the energy range $(0,1.3)$ GeV. The purple region denotes the ``experimental'' Adler function from tau data~\cite{Davier:2005xq}. The black line represents the Adler function as in Fig.~\ref{Adler}. For a slightly different value of the constants $C,K,c_1$, the dashed, red line represents the Adler function saturating the muon $g-2$ discrepancy between experiments and predictions. The inset is a zoom on the region of interest. }
\label{TabAdleramu}
\end{figure}

We consider now the so-called hadronic vacuum polarization (h.v.p.) contribution to the magnetic moment of the muon $\vec{\mu}$ -- for analyses on this subject see Refs.~\cite{Czarnecki:2000id,Czarnecki:2001pv,Knecht:2003kc,Davier:2004gb,Miller:2007kk,Jegerlehner:2009ry,Miller:2012opa,Aoyama:2020ynm,Cvetic:2020naz}.
It is given by
\begin{equation}
\vec{\mu}= g \frac{Q_e}{2m_{\mu}c}\vec{s}\,,
\end{equation}
where $\vec{s}$ is the spin, $Q_e$ is the electric charge, $m_{\mu}$ is the muon mass, $c$ is the speed of light, and Dirac's theory predicts $g = 2$. Quantum effects correct the value $g=2$ and the deviation is parameterized as $a_{\mu}=(g-2)/2$. Comprehensive analyses of muon $g-2$ within the Standard Model can be found in Ref.~\cite{Davier:2010nc,Davier:2017zfy,Davier:2019can,Aoyama:2020ynm}.

The leading order hadronic vacuum polarization contribution in terms of the QCD Adler function is of the form~\cite{Lautrup:1971jf,Knecht:2003kc}
\begin{equation}\label{amu_Adler}
a_{\mu}^{(\text {h.v.p.} )}=2 \pi^{2}\left(\frac{\alpha}{\pi}\right)^{2} \int_{0}^{1} \frac{d x}{x}(1-x)(2-x) D\left(Q\right)\,,
\end{equation}
where $\alpha\simeq 1/137$ is the electromagnetic coupling constant, and $Q = \sqrt{\frac{x^{2}}{1-x} m_{\mu}^{2}}$.

Although the numerical instabilities for the parameters discussed in the previous section, it is worth asking
whether our model can implement the \emph{tentative idea} proposed in Ref.~\cite{Keshavarzi:2020bfy}. The authors studied the possibility that the $g-2$ discrepancy
could be solely explained by modifying the SM vacuum polarization function contribution, deeming this scenario rather unlikely. A modification of the h.v.p. can be in tension with electro-weak precision tests~\cite{PhysRevLett.125.091801,Malaescu:2020zuc}~\footnote{Constraints on the h.v.p. are also discussed in Ref.~\cite{Colangelo:2020lcg}. Direct measurement of the h.v.p. will definitively shed light on the subject~\cite{Banerjee:2020tdt}.} Although improbable, the authors of Ref.~\cite{Keshavarzi:2020bfy} still noted that there might be a missed contribution for $Q\lesssim 0.7$ GeV, an energy range in which constraints do not yet rule out the possibility of explaining the $g-2$ discrepancy by deviations of the $e^+ e^-$ cross-section measurement. Interestingly, this would be consistent with the most recent lattice evaluation~\cite{Borsanyi:2020mff}.

We wish to consider the impact of this hypothesis on our model.
Thus we require the Adler function to match the experimental data for energies $Q\geq 0.7$ GeV and instead to allow for some deviations below $Q< 0.7$ GeV.
\begin{table}
\centering
\tabcolsep=0.05cm
\begin{tabular}{ r | c }
\hline
 \hline
  Parameter & $a_{\mu}$ discrepancy \\
\hline
\hline
$K$ & 0.865 \\
$C$ & 0.764 \\
$c_1$ & -0.184 \\
 \hline
$\Lambda$ & 677 MeV \\
 \hline \hline
\end{tabular}
\caption{Values reproducing the experimental $g-2$ discrepancy.}
\label{amufit}
\end{table}
The integral in Eq.~\eqref{amu_Adler} requires the Adler function $D(Q)$ in the energy range $[0,\infty)$.
Following Ref.~\cite{Peris:1998nj}, one has to split $D$ in two branches, using the perturbative estimate for $Q>\sqrt{1.6}$ GeV and the data for $\leqslant\sqrt{1.6}$ GeV. The Eq.~\eqref{trans_adler} provides a good estimate of data. Thus the Adler function used in the evaluation of Eq.~\eqref{amu_Adler} is given by
\begin{equation}\label{pwAdler}
D(Q)=\left\{
\begin{array}{ll}
 D_{resurg.}(Q) & Q\leqslant\sqrt{1.6} \text{ GeV} \\
 D_{pert.}(Q) & Q>\sqrt{1.6} \text{ GeV} \,.
\end{array}
\right.
\end{equation}
The corresponding behavior of the Adler function in $[0,1.3]$ GeV is shown in Fig.~\ref{TabAdleramu} (solid black line), together with the experimental uncertainties represented by the light blue band.

The modification of the value of the constants $K$, $c_1$ and $C$, shown in Tab.~\ref{amufit}, saturates the gap between the average value of $a_\mu^{(\text {h.v.p.} )}\simeq6.9\times 10^{-8}$~\cite{Miller:2007kk,Miller:2012opa} and the value $a_\mu^{(\text {h.v.p.} )}\simeq7.15\times 10^{-8}$ consistent with the Muon $g-2$ Collaboration~\cite{PhysRevLett.126.141801} experimental result. The deviations with respect the values in Tab.~\ref{tabrersurgencefit} (third column) are about few percent for $K$ and $\sim100\%$ for $C$ and $c_1$. Notice that the deviations are outside the range estimated in the previous section varying the starting perturbative information. This is expected since, by construction, we are now describing a modified Adler function, such that it is in agreement with $a_\mu$ measurement and no longer the current "experimental" one in Fig.~\ref{Adler}. The plot for the Adler function, corresponding to the values for $C,c_1$ and $K$ in Tab.~\ref{amufit}, is shown in Fig.~\ref{TabAdleramu} represented with the dashed red line.

In our picture the deviation concerning the average value $a_{\mu}\simeq6.9\times 10^{-8}$ of Refs.~\cite{Miller:2007kk,Miller:2012opa} is due to non-perturbative (non-analytic) contributions in the strong coupling constant $\alpha_s$, which were calculated using the resurgence framework of Refs~\cite{Bersini:2019axn,Maiezza:2019dht}. These non-analytic contributions become dominant for $\alpha_s\sim1$. The nonperturbative electro-weak corrections are sub-leading since the numerical values of the electromagnetic and weak couplings remain small at the muon mass-energy scale.

\section{Summary and outlook}

We have shown that the analytical expression for the Adler function in Eq.~\eqref{trans_adler} has the flexibility to reproduce the Adler function data at the hadronic scale within the range of current determinations. We have used Cornwall's coupling to model the running of $\alpha_s$ whose value freezes at low energies with $\alpha_s(0) \sim \mathcal{O}(1)$. In this work, we focused on the latter effective description for the running of $\alpha_s$ due to its simplicity. Cornwall's coupling ensures the applicability of the resurgent approach to renormalons and renormalization group equation of Ref.~\cite{Bersini:2019axn}, which relies on non-linear, ordinary differential equations~\cite{CostinBook}. As a result, Eq.~\eqref{trans_adler} features 3+1 arbitrary parameters ($K, C, c_1$ and $\Lambda$), in contrast to conventional renormalon-based evaluations with an infinite number of arbitrary constants.

We have determined those parameters from data, shown in Tab.~\ref{tabrersurgencefit}, and our representation of the Adler function is drawn in Fig.~\ref{Adler}. At the present level, our method is not yet quantitatively stable; namely, the values of the fitted parameters are sensitive to the perturbative information that one starts with -- as expected.
There is the possibility that the knowledge of higher-loop corrections for the Adler function would stabilize the values for the parameter shown in Tab.~\ref{tabrersurgencefit}. However, one potential problem against this possibility is that the coefficients $\delta_n$ in Eq.~\eqref{Adlerpert} also receive $n!$ contributions independent from the renormalons due to the instantons, related to the proliferation of the number of Feynman diagrams as the order of perturbation theory $\mathcal{O}(\alpha_s^n)$ increases. This means there are two superimposed Stokes lines, one due to the renormalons and the other due to the instantons. This situation is often called ``resonance'' in the mathematical literature. The treatment of renormalons, proposed here and the previous Refs.~\cite{Bersini:2019axn,Maiezza:2021mry}, is in the approximation of non resonance, as also assumed in Ref.~\cite{CostinBook}. Giving up on this assumption leads to a complex mathematical challenge of considering the instantons on top of the renormalons, and to our knowledge, this is an open mathematical problem. However, since renormalons induce the nearest singularity from the origin of the Borel plane, they dominate the large order behavior of the perturbative expansion~\cite{Parisi:1978iq} as long as the coupling $\alpha_s\sim1$. We speculate that the non-resonant effects from instanton and renormalon singularities might give a subleading contribution that would not drastically modify our conclusions.

Notwithstanding the above sources of theoretical uncertainties, the IR description of the Adler function provided by resurgence gives a remarkable improvement to standard analytical approaches -- with an infinite number of arbitrary constants. Still, we have not addressed the impact on the $K, C, c_1$ and $\Lambda$ from the inclusion of the chiral symmetry-breaking effects due to quark masses. However, we expect the uncertainties due to the perturbative inputs in Tab.~\ref{tabrersurgencefit} to be the dominant ones, and these issues are left for future work. The interplay with different processes would open the possibility of testing the universality of QCD running coupling and of the constants in Tab.~\ref{tabrersurgencefit}, in the spirit of Ref.~\cite{Dokshitzer:1995qm}. Indeed, we expect our result to apply to other relevant processes involving the two-point Green function at the hadronic scale. An example may be the event shape observables in $e^+e^-$ collisions~\cite{Webber:1994cp,Manohar:1994kq,Korchemsky:1994is,Dokshitzer:1995zt,Akhoury:1995sp,Nason:1995np}. Other applications may be on the determination of the heavy quark pole mass~\cite{Beneke:1994sw,Bigi:1994em} and on the static quark-antiquark potential~\cite{Aglietti:1995tg}.

We have also addressed the implications and the interplay with the muon's magnetic moment. In particular, we implemented in our model the \emph{tentative hypothesis} to explain the SM discrepancy for $a_\mu$~\cite{PhysRevLett.126.141801} by modifying the vacuum-hadronic-polarization contribution. As shown in Fig.~\ref{TabAdleramu}, its only effect is slightly spoiling the behavior of the Adler function at energies $\lesssim0.7$ GeV, a range in which data may not be complete due to missed contributions in the hadronic cross-section $\sigma(e^+e^-\rightarrow hadrons)$. The corresponding modification of the (fitted) parameters that would explain the $g-2$ discrepancy is shown in Tab.~\ref{amufit}. As proof of concept, our result implies that the muon $g-2$ discrepancy can be accounted for by including non-analytic contributions in the strong coupling constant $\alpha_s$ -- calculated using resurgence theory. The latter would be compatible with the most recent lattice calculation~\cite{Borsanyi:2020mff}.

%%%%%%%%%%%%%%%%%%%%%%%%%%%%%%%%%%%%%%%%%%%%%%%%%%%%%%%%%%%%%%%%%%%%%%%%%%%%%%%%%%%%%%%%%%%%%%%%%%%%%%%%%%%%%%%%%%%%%%%%%%%%%%%%%%%%%%%%%%%%%%%%%%%%%%%%%%%%%%%%%%%%%%%%%%%%%%%%%%
\section*{Acknowledgements}
AM dedicates this work to the memory of his father, Agostino.
JCV was supported in part under the U.S. Department of Energy contract DE-SC0015376.

\appendix

%%%%
%%%%%%
\section{Borel-Ecalle resummation based on the non-linear Ordinary Differential Equations}\label{app:Synthesis}

In Chapter 5 of Ref.~\cite{CostinBook}, the author developed a generalization of the Borel resummation called  Borel-Ecalle resummation (or  "synthesis"). This generalized resummation procedure is based on non-linear  ODEs, so one can resum $n!$ divergent series that are otherwise non-Borel summable.
We summarize the main points necessary for the Borel-Ecalle resummation of the renormalon singularities in QFT. To this end, consider the generic first-order ODE
\begin{equation}\label{gen_ODE}
y'=f(x,y(x))\,,
\end{equation}
and formally expand the function $f$ for large $x$ and small $y$~\cite{CostinBook}
\begin{equation}\label{ODE_x}
y(x)'= f_0(x)-\zeta y(x) +\frac{1}{x}a_p y(x) + g(x,y(x))\,,
\end{equation}
where $f_0(x)$ is an analytic function of $x$,  $g$ is an analytic function at $(0,\infty)$ and\\
\noindent
$g=\mathcal{O}(x^{-2},y(x)^2,x^{-2}y(x))$. The formal power series solution $y_0(x)= b_n x^{-n}$  is general a divergent series, namely $b_n\sim n!$. To overcome this difficulty,   one considers its associated formal transseries solution:
\begin{equation}
y(x) = \sum_{k=0}^{\infty} C^k x^{a_p \,k}e^{-k\zeta/x}y_k(x)\,.\label{trasseries_formal}
\end{equation}
where $C$ is a real constant provided the coefficients $b_n$ are all real. As shown in Ref~\cite{CostinBook}, the Borel transform $Y_0(z)$ associated to $y_0(x)$ has an infinite number of singularities located at  $k\,\zeta $. Expanding in a neighborhood of a given singularity $n\zeta$ for fixed $n$,  the Borel transform $Y_0(x)$ have the following analytic structure
\begin{equation}
\mathbf{Y}_{0}(p)\propto \left\{\begin{array}{l} \Theta(z-k\zeta)(z-k\zeta)^{-1-a_p}+...\,,a_p \neq -1 \\ \\
\Theta(z-k\zeta)\log(z-k\zeta)+...\,, a_p = -1\end{array}\right.
\end{equation}
where $k$ takes integer values between $[0,\infty)$ and the parameters $\zeta,a_p$ are the same ones entering in the transseries solution in Eq.~\eqref{trasseries_formal}.  From the preceding discussion, we can see that both $\zeta,a_p$ are the most relevant parameters since they dictate the position of the singularities and the analytic structure of the Borel transform of the solution $y(x)$, respectively.

Next, consider the Borel transforms $Y_k(z)$ of the functions $y_n(x)$ in Eq.~\eqref{trasseries_formal}.   In Ref.~\cite{CostinBook}, it is shown that the singularities of $Y_k(z)$ (for $k\geq 1$) are also those  in $Y_0(z)$, and as will shall see this later point becomes apparent from the resurgence expressions relating $Y_0(z)$ with $Y_k(z)$.   For each function $Y_k(z)$ $\forall k \in \mathbb{N}+ \{0\}$  one builds the functions $Y^{\pm}_k(z) \equiv Y_k(z\pm i \epsilon)$, which are nothing but the analytic continuation of the functions $Y_k(z)$ above and below the positive real axis, respectively.

Once $Y_0(z)$ is known (to all orders),  \emph{resurgence} is the property that allows the functions $Y_k$  to  be written in terms of $Y_0$ using the following operation:
\begin{equation} \label{eq:resurgence}
i^k S^k Y_k=(Y_0^- - Y_0^{-k-1+}) \circ \tau_k, \quad \tau_k: z \mapsto z +k\zeta\,,
\end{equation}
where $S$ is the nonperturbative Stokes  constant and
\begin{equation}\label{eq:analyticcontinuation}
Y_k^{-m+}  = Y_k^+ + \sum_{j=1}^m  {k+j \choose k} S^j  Y^+_{k+j}\circ \tau_{-j}\,.
\end{equation}
One Stokes constant is associated with each singular direction in the Borel plane. For $a_p\in \mathbb{N}+ \{0\}$  from Eq.~\eqref{trasseries_formal} and Eq.~\eqref{eq:resurgence} one can prove that only the combination $C/S$ enters in the expression for $y(x)$, which means the constant $C/S$ can be then treated as a single arbitrary constant (that we denote again as $C$ for simplicity). It is worth commenting that for renormalons, $C$ cannot be determined from first principles and must be fixed from data. This work deals with one singular direction with singularities in the real axis at $z_{pole}= \zeta,  2\zeta, 3\zeta,...$.

Next, one can construct  the \emph{balanced average} associated to each $Y_k$
\begin{equation}\label{balanced:average}
Y_k^{bal} \equiv Y_k^+ + \sum_{n=1}^{\infty}2^{-n} (Y_k^- - Y_k^{-n-1+})\,.
\end{equation}
This definition guarantees that when  $y_0$ is a real formal series, then $Y_k^{bal}$ is also real. In this case, the formula can be symmetrized by taking $1/2$ of the above expression plus $1/2$ of the same expression with $+$ and $-$ interchanged~\cite{costin1998,CostinBook}  (see proposition (5.77) and Eq.~(5.118) of Ref.~\cite{CostinBook}). The balanced average preserves all the algebraic operations, differentiation, integration, function compositions, and convolutions.

Finally, one performs the Laplace transform along a Stokes direction. As discussed,  when the $Y_k(z)$ has poles in the positive real axis, the Borel-Laplace resummation is generalized to the Borel-Ecalle resummation (denoted as $\mathcal{E}$)  discussed  above, and  the final resummed result for each $Y_k$ function  is given by
\begin{equation}
\mathcal{E}(y_k)= \mathcal{L \circ B}(y_k) = \mathcal{L}(Y_k) = \int_0^{\infty} Y_k^{bal} e^{-z/x}dz\,,
\end{equation}
Such that the actual result of summing the original series $y_0(x)$ is
\begin{equation}\label{resummedy}
y_0(x) \mapsto y(x) = \mathcal{E}(y_0)(x) + \sum_{k=1}^{\infty} e^{-k\zeta /x} x^{a_p k}\mathcal{E}(y_k) (x)\,,
\end{equation}
When no poles are present in the positive real axis, the usual Borel-Laplace resummation procedure is recovered.

\section{Highlights on ODE-based resurgence}\label{ODE_RES}

We briefly summarize the points on which the ODE-based resummation of renormalons is built, based on Refs.~\cite{Bersini:2019axn,Maiezza:2021mry}, in which the mathematical
theory highlighted in the previous appendix was connected to RGE.

Consider the two-point correlator
$$
\int d^{4} x \mathrm{e}^{-i q x}\left\langle 0\left|T\left(A_{\mu}(x) A_{\nu}(0)\right)\right| 0\right\rangle=\mathrm{i}\left(q_{\mu} q_{\nu}-q^{2} g_{\mu \nu}\right) G\left(Q^{2}\right)
$$
where $Q^2=-q^2$. The function $G(L, \alpha_s)$ satisfies RGE:
\begin{equation}\label{RGE}
\left[-\partial_{L}+\beta(\alpha_s) \partial_{\alpha_s}- \gamma(\alpha_s)\right] G(L, \alpha_s)=0\,,
\end{equation}
where $L=\ln(\frac{Q^2}{\mu^2})$. In full generality, one can write $G$ as its perturbative part plus a genuine nonperturbative function $R$:
\begin{equation}\label{definition}
G(\alpha_s) : = \sum_{i=0}^{\infty} \gamma_i(\alpha_s)L^i + R(\alpha_s)\,.
\end{equation}
The function $R$ contains \emph{by definition} all the $n!$ contributions due to the renormalon diagrams. If one knew $\beta$ and $\gamma$ in Eq.~\eqref{RGE}, $G$ would be completely determined. Thus one concludes that if $G$ is a function of $R$,  both $\beta$ and $\gamma$ must also be a function of $R$.

Pugging Eq.~\eqref{definition} into Eq.~\eqref{RGE} and expanding for small $\alpha_s$ and $R$, then using $\gamma =\gamma_1 +qR$+...,  one gets  for $R(\alpha_s)$ the same type of  nonlinear ODE studied in Ref.~\cite{Bersini:2019axn}, which  is in the form of Eq.~\eqref{ODE_x}, provided the change of variable $x=1/\alpha_s$ is made, namely
\begin{equation}\label{main}
\frac{d R(\alpha_s)}{d \alpha_s} =\frac{ q}{\beta_0 \alpha_s^2} R(\alpha_s) -a_p\frac{R(\alpha_s)}{\alpha_s}
+ a_h+\mathcal{O}(R(\alpha_s)^2)\,
\end{equation}
where $a_{p,h}$ are  functions of the coefficients of asymptotic expressions of $\beta,\gamma$. Notice that  $\zeta= q/\beta_0$.

The expansion in $R$ is formal, in the same sense of Eqs.~\eqref{gen_ODE} and~\eqref{ODE_x}. In other words, one can consider any power of $R$ in Eq.~\eqref{main}, but the specific form of the nonlinearity in $R$ is irrelevant for the approximate solution in Eq.~\eqref{trasseries_formal}. Conversely, the presence of the nonlinearities determines the \emph{fundamental property}  of the equation, namely its solution in Borel space ($\alpha_s\mapsto z$) features an infinite number of poles in $z=q/\beta_0$.

Finally, by matching with the skeleton diagram evaluation in an asymptotically free model (i.e., QCD), one can identify $R$ with the resummed IR renormalons, setting $q=-1$. $a_p$ determines the kind of poles and, in this work, we are interested in $a_p=1$, corresponding to quadratic poles, which are the ones emerging from the direct computations~\cite{Neubert:1994vb} (see the last appendix).

Transseries solution of Eq.~\eqref{main} is
\begin{equation}\label{transseries}
R(\alpha_s) = \sum_{n=0}^{\infty}C^n R_n(\alpha_s)\,\alpha_s^{-n\,a_p}\,e^{-\frac{n\,q}{\beta_0\alpha_s}}\,,
\end{equation}
where $R_0$ is the series solution of Eq.~\eqref{main}, and all the $R_n$ are calculable recursively from $R_0$ by Eq.~\eqref{eq:resurgence}. The bottom line is that Eq.~\eqref{main} is first-order ODE with thus one arbitrary parameter $C$ in Eq.~\eqref{transseries}. Therefore, IR renormalons can be resummed, and the result is determined up to only one arbitrary parameter.

\section{Resurgence and the Adler function}\label{Detail_RRGE_Adler}

In this appendix, we present the expressions necessary to derive the transseries for the Adler function in Eq.~\eqref{trans_adler}, using the formalism of Ref.~\cite{CostinBook}. The  renormalon contribution to the Adler function is given by~\cite{Neubert:1994vb}
\begin{align}\label{rearrange}
\frac{1}{C_F\,K}B[&D_{bubble}](z) = \frac{3 e^{10/3} \mu ^4}{2 \beta_0Q^4 \left(\frac{2}{\beta_0}+z\right)}
+ \frac{e^5 \mu ^6 \left(6 \log \left(\frac{\mu
 ^2}{Q^2}\right)+1\right)}{6\beta_0 Q^6
 \left(\frac{3}{\beta_0}+z\right)} -\frac{e^5 \mu^6}{\beta_0^2 Q^6 \left(\frac{3}{\beta_0}+z\right)^2} -\nonumber\\
& \sum_{p=1}^{\infty}\left( \frac{ \mu ^4 e^{\frac{10 p}{3}+\frac{10}{3}}
 \left(\frac{Q}{\mu }\right)^{-4 p} \left(12 p^2 \log
 \left(\frac{\mu ^2}{Q^2}\right)+20 p^2+6 p \log \left(\frac{\mu
 ^2}{Q^2}\right)-2 p-3\right)}{6 \beta_0p^2 (2 p+1)^2 Q^4
 \left(\frac{2 p+2}{\beta_0}+z\right)} + \right. \nonumber \\
 & \frac{ \mu ^6 e^{\frac{10 p}{3}+5} \left(\frac{Q}{\mu
 }\right)^{-4 p} \left(12 p^2 \log \left(\frac{\mu
 ^2}{Q^2}\right)+20 p^2+18 p \log \left(\frac{\mu
 ^2}{Q^2}\right)+18 p+6 \log \left(\frac{\mu
 ^2}{Q^2}\right)+1\right)}{6 \beta_0(p+1)^2 (2 p+1)^2 Q^6
 \left(\frac{2 p+3}{\beta_0}+z\right)}\nonumber \\
 & \left. -\frac{ \mu ^6 e^{\frac{10 p}{3}+5} \left(\frac{Q}{\mu}\right)^{-4 p}}{\beta_0^2 (p+1) (2 p+1) Q^6
 \left(\frac{2 p+3}{\beta_0}+z\right)^2}+\frac{
 \mu ^4 e^{\frac{10 (p+1)}{3}} \left(\frac{Q}{\mu }\right)^{-4
 p}}{\beta_0^2 p (2 p+1) Q^4 \left(\frac{2
 p+2}{\beta_0}+z\right)^2}\right)\,, \nonumber \\
 \end{align}
where $C_F=\frac{4}{3}$ $K$ is an overall, arbitrary constant of the large order behavior. The Eq.~\eqref{rearrange} does not contain UV renormalon contribution. We should recall that these do not lead to ambiguities since they are outside the path of integration in the Laplace transform. Moreover, we find the UV contribution negligibly small in the considered energy regime (below 2.5 GeV), and then we omit it.

As said, we have to consider the leading poles, i.e., the quadratic ones, in the Borel-Ecalle resummation of Eq.~\eqref{rearrange} and, separately, the simple pole at $z=\frac{2}{\beta_0}$.
The solution of Eq.~\eqref{main} is then found at the leading order~\cite{CostinBook} from the leading \emph{quadratic} infinite string of poles in the Borel transform
\begin{align}\label{Borel_resurgence}
\frac{1}{K\,C_F}B[&D_{bubble}](z) \rightarrow \frac{3 e^{10/3} \mu ^4}{2 \beta_0Q^4 \left(\frac{2}{\beta_0}+z\right)}
 -\frac{e^5 \mu^6}{\beta_0^2 Q^6 \left(\frac{3}{\beta_0}+z\right)^2} -\nonumber\\
 & \sum_{p=1}^{\infty} \left[ \frac{
 \mu ^4 e^{\frac{10 (p+1)}{3}} \left(\frac{Q}{\mu }\right)^{-4
 p}}{\beta_0^2 p (2 p+1) Q^4 \left(\frac{2
 p+2}{\beta_0}+z\right)^2} -\frac{ \mu ^6 e^{\frac{10 p}{3}+5} \left(\frac{Q}{\mu}\right)^{-4 p}}{\beta_0^2 (p+1) (2 p+1) Q^6
 \left(\frac{2 p+3}{\beta_0}+z\right)^2} \right]\,. \nonumber \\
 \end{align}
A convenient choice of renormalization scale we adopt is $\mu^2= Q^2e^{-5/3}$~\cite{Broadhurst:1992si,Neubert:1994vb}.  The estimate of the fermion-bubble diagram contribution to the QCD Adler function~\cite{Neubert:1994vb} gives infinite quadratic poles starting at $z=-3/\beta_0$. By means of Eq.~\eqref{main} and the previous discussion, we have to set $a_p =1+ \mathcal{O}(\beta_1/\beta_0^2)$(so we neglect the two-loop corrections proportional to $\beta_1$) and identify $R$ with the Borel-Ecalle resummation of the quadratic poles. The simple remnant pole at $z=- 2/\beta_0$ cannot be incorporated into the generalized resummation. Thus we parameterize its associated ambiguity (which we name $c_1$) consistently with
\begin{equation}\label{ambiguity}
\left(z+\frac{2}{\beta_0}\right)^{-1} \mapsto -2\pi c_1\,  e^{\frac{2}{\beta_0\alpha_s}}\,.
\end{equation}

\subsection{Quadratic poles}

We now Borel-Ecalle resum the quadratic poles in Eq.~\eqref{Borel_resurgence}. For the sake of illustration, first, consider the simple example of the Borel transform $Y(z)$ of a given function $y(\alpha_s)$
\begin{equation}
Y(z)=\sum_n (z+n)^{-2}\,,
\end{equation}
Furthermore, apply the material quoted in appendix~\ref{app:Synthesis}.
Eq.~\eqref{eq:resurgence} reduces to
\begin{equation}\label{example0}
Y_1(z)\propto (Y^-(z)-Y^+(z))\circ \tau_1 = - 2  \pi  \sum_n \Theta(z-1+n) \delta'(z-1+n)\,,
\end{equation}
and all the $Y_n=0$ if $n>1$.
Eq.~\eqref{resummedy} reduces to
\begin{equation}\label{example1}
y(\alpha_s)=y_0(\alpha_s)+C \frac{1}{\alpha_s}e^{\frac{1}{\alpha_s}} y_1 (\alpha_s)\,,
\end{equation}
with
\begin{align}\label{example2}
&y_0(\alpha_s)= \sum_n \mathcal{L}_{PV}[(z+n)^{-2}] = \sum_n \left[\frac{1}{n}-e^{\frac{n}{\alpha_s}}\alpha_s \Gamma(0,\frac{n}{\alpha_s})\right]\,, \nonumber \\
&y_1(\alpha_s)=-2 \pi \sum_n \frac{e^{\frac{n-1}{ \alpha_s}}}{\alpha_s}\,,
\end{align}
being $\mathcal{L}_{PV}$ the Cauchy principal value of the Laplace integral and $\Gamma(0,\frac{n}{\alpha_s})$ the incomplete Gamma function, and $y_1(\alpha_s)$ the Laplace
transform of Eq.~\eqref{example0}.

It is now sufficient to apply the same logic of Eqs.~\eqref{example1},~\eqref{example2} for the quadratic poles in Eq.~\eqref{Borel_resurgence}.
As a matter of fact, one takes the principal value of Eq.~\eqref{Borel_resurgence} to get $D_K$ in Eq.~\eqref{trans_adler}, while one does the replacement
\begin{equation}
(z +\frac{n}{\beta_0})^{-2} \mapsto - 2 \pi \frac{1}{\alpha_s} e^{\frac{n-1}{\beta_0 \alpha_s}}
\end{equation}
in Eq.~\eqref{Borel_resurgence} to get $D_1$ in Eq.~\eqref{trans_adler}. Putting together,
one gets
\begin{align}\label{trans_adler_app}
D_{resurg.}(Q) =\, & D_0(Q)-\frac{4\pi}{\beta_0} c_1e^{\frac{2}{\beta_0\,\alpha_s(Q^2)}}\nonumber \\
& + C e^{\frac{1}{\beta_0\,\alpha_s(Q^2)}}\left(\frac{1}{\alpha_s(Q^2)}\right)^{a_p}D_1(Q^2)\,,
\end{align}
where the part in $c_1$ is related to the simple pole parameterized by Eq.~\eqref{ambiguity}, and $D_0(Q)$ contains the perturbative expression up to $\mathcal{O}(\alpha_s^4)$ and the higher-order $n!$ corrections due to the fermion-bubble diagrams. The $n!$ part is regularized by the Cauchy principal value of the Laplace integral of Eq.~\eqref{Borel_resurgence} and is given by:
\begin{equation}
D_0(Q) = D_{pert}(Q)+  D_{K}(Q)\,.
\end{equation}
The function  $D_{pert}(Q)$ is the expression of the Adler function found in perturbation theory,
\begin{equation}
D_{pert}\left(Q\right)=1+\frac{\alpha_{s}}{\pi} \sum_{n=0}^{3} \alpha_{s}^{n}\left[d_{n}\left(-\beta_{0}\right)^{n}+\delta_{n}\right]\,.
\end{equation}
The $n!$ contribution is proportional to the unknown constant $K$, which parameterizes that one is summing a few perturbative terms with no $n!$ behavior to the renormalons series that approximates the large order ($n!$) behavior. Let us recall that $K$ cannot be determined \textit{a priori}. Since renormalons do not have a semiclassical limit, it is impossible to estimate which order in perturbation theory evaluations starts behaving factorially. This issue is also related to the fact that one cannot determine an optimal truncation for the renormalon series since doing so would require using semiclassical methods.
Summing over $p$ Eq.~\eqref{Adler_K} gives
\begin{align}
& \frac{D_{K}(Q)}{K} \simeq \frac{4 e^{\frac{3}{\text{$\beta_0 $} \alpha _s}} \Gamma
   \left(0,\frac{3}{\text{$\beta_0 $} \alpha _s}\right)}{3 \text{$\beta_0$}^2
   \alpha _s}+\frac{2 e^{\frac{2}{\text{$\beta_0$} \alpha_s}}
   \Gamma \left(0,\frac{2}{\text{$\beta_0$} \alpha _s}\right)}{\text{$\beta_0$}}  \nonumber \\
   &+\frac{601 \text{$\beta_0$}^4 \alpha _s^4-390 \text{$\beta_0$}^3
   \alpha_s^3+390 \text{$\beta_0$}^2 \alpha _s^2-828 \text{$\beta_0$} \alpha
   _s-432}{972 \text{$\beta_0$}}
\end{align}
This is the equivalent of the first of the Eqs.~\eqref{example2}.
Products of the type $4 e^{\frac{3}{\text{$\beta_0 $} \alpha _s}} \Gamma\left(0,\frac{3}{\text{$\beta_0 $} \alpha _s}\right)$ are Taylor expandable and hence contains all powers in $\alpha_s$. The terms up to $\mathcal{O}(\alpha^4_s)$ are subtracted in agreement with the known low order contributions in $D_{pert}(Q)$.
Finally
\begin{align}\label{nonpert_adler}
D_1(Q) = & \frac{8 \pi K }{3 \alpha_s \beta_0^2}\left[2e^{\frac{1}{\alpha_s\beta_0}} -
 \left(e^{\frac{1}{\alpha_s\beta_0}}+1\right)
 \log \left(1-e^{\frac{2}{\alpha_s \text{$\beta
 $0}}}\right)\right. \nonumber \\
 &-\left. 2 \left(e^{\frac{1}{\alpha_s \text{$\beta
 $0}}}+1\right) \tanh ^{-1}\left(e^{\frac{1}{\alpha_s
 \beta_0}}\right)\right]\,.
\end{align}
This is the equivalent of the second of the Eqs.~\eqref{example2}.
Considering $\beta_1$ does not change the position of the poles, but it would modify the analytic structure of the Borel transform. Including the two-loop coefficient $\beta_1$ would only give subleading corrections $\mathcal{O}(\beta_1/\beta_0^2)$. In particular, one would have
higher powers of $C$ suppressed by $\beta_1/\beta_0^2$ (multiple nonperturbative sectors~\cite{CostinBook}) in the example of Eq.~\eqref{example1} and the  Adler function case in Eq.~\eqref{trans_adler_app}.

\bibliographystyle{jhep}
\bibliography{biblio}

\providecommand{\href}[2]{#2}\begingroup\raggedright\begin{thebibliography}{10}

\bibitem{PhysRevD.10.3714}
S.~L. Adler, \emph{Some simple vacuum-polarization phenomenology:
  ${e}^{+}{e}^{\ensuremath{-}}\ensuremath{\rightarrow}\mathrm{hadrons}$; the
  muonic-atom x-ray discrepancy and
  ${\mathrm{g}}_{\ensuremath{\mu}}\ensuremath{-}2$},
  \href{http://dx.doi.org/10.1103/PhysRevD.10.3714}{\emph{Phys. Rev. D} {\bf
  10} (Dec, 1974) 3714--3728}.

\bibitem{Baikov:2008jh}
P.~A. Baikov, K.~G. Chetyrkin and J.~H. Kuhn, \emph{{Order alpha**4(s) QCD
  Corrections to Z and tau Decays}},
  \href{http://dx.doi.org/10.1103/PhysRevLett.101.012002}{\emph{Phys. Rev.
  Lett.} {\bf 101} (2008) 012002}, [\href{http://arxiv.org/abs/0801.1821}{{\tt
  0801.1821}}].

\bibitem{Francis_2013}
A.~Francis, B.~Jäger, H.~B. Meyer and H.~Wittig, \emph{New representation of
  the adler function for lattice qcd},
  \href{http://dx.doi.org/10.1103/physrevd.88.054502}{\emph{Physical Review D}
  {\bf 88} (Sep, 2013) }.

\bibitem{PhysRevD.10.3235}
D.~J. Gross and A.~Neveu, \emph{Dynamical symmetry breaking in asymptotically
  free field theories},
  \href{http://dx.doi.org/10.1103/PhysRevD.10.3235}{\emph{Phys. Rev. D} {\bf
  10} (Nov, 1974) 3235--3253}.

\bibitem{LAUTRUP1977109}
B.~Lautrup, \emph{On high order estimates in qed},
  \href{http://dx.doi.org/https://doi.org/10.1016/0370-2693(77)90145-9}{\emph{Physics
  Letters B} {\bf 69} (1977) 109--111}.

\bibitem{tHooft:1977xjm}
G.~'t~Hooft, \emph{{Can We Make Sense Out of Quantum Chromodynamics?}},
  {\emph{Subnucl. Ser.} {\bf 15} (1979) 943}.

\bibitem{Parisi:1978iq}
G.~Parisi, \emph{{The Borel Transform and the Renormalization Group}},
  \href{http://dx.doi.org/10.1016/0370-1573(79)90111-X}{\emph{Phys.\ Rept.}
  {\bf 49} (1979) 215--219}.

\bibitem{wilson1972}
K.~G. Wilson and W.~Zimmermann, \emph{Operator product expansions and composite
  field operators in the general framework of quantum field theory},
  {\emph{Comm. Math. Phys.} {\bf 24} (1972) 87--106}.

\bibitem{Shifman:2021iis}
M.~Shifman, \emph{{Yang-Mills at Strong vs. Weak Coupling: Renormalons, OPE And
  All That}},  \href{http://arxiv.org/abs/2107.12287}{{\tt 2107.12287}}.

\bibitem{Beneke:1998ui}
M.~Beneke, \emph{{Renormalons}},
  \href{http://dx.doi.org/10.1016/S0370-1573(98)00130-6}{\emph{Phys. Rept.}
  {\bf 317} (1999) 1--142}, [\href{http://arxiv.org/abs/hep-ph/9807443}{{\tt
  hep-ph/9807443}}].

\bibitem{Shifman:2013uka}
M.~Shifman, \emph{{New and Old about Renormalons: in Memoriam Kolya Uraltsev}},
  \href{http://dx.doi.org/10.1142/S0217751X15430010}{\emph{Int. J. Mod. Phys.
  A} {\bf 30} (2015) 1543001}, [\href{http://arxiv.org/abs/1310.1966}{{\tt
  1310.1966}}].

\bibitem{Cvetic:2018qxs}
G.~Cveti\v{c}, \emph{{Renormalon-motivated evaluation of QCD observables}},
  \href{http://dx.doi.org/10.1103/PhysRevD.99.014028}{\emph{Phys. Rev. D} {\bf
  99} (2019) 014028}, [\href{http://arxiv.org/abs/1812.01580}{{\tt
  1812.01580}}].

\bibitem{Caprini:2020lff}
I.~Caprini, \emph{{Conformal mapping of the Borel plane: going beyond
  perturbative QCD}},
  \href{http://dx.doi.org/10.1103/PhysRevD.102.054017}{\emph{Phys. Rev. D} {\bf
  102} (2020) 054017}, [\href{http://arxiv.org/abs/2006.16605}{{\tt
  2006.16605}}].

\bibitem{Shirkov:1997wi}
D.~V. Shirkov and I.~L. Solovtsov, \emph{{Analytic model for the QCD running
  coupling with universal alpha-s (0) value}},
  \href{http://dx.doi.org/10.1103/PhysRevLett.79.1209}{\emph{Phys. Rev. Lett.}
  {\bf 79} (1997) 1209--1212}, [\href{http://arxiv.org/abs/hep-ph/9704333}{{\tt
  hep-ph/9704333}}].

\bibitem{Nesterenko:2007fm}
A.~V. Nesterenko, \emph{{Adler function in the analytic approach to QCD}},
  {\emph{eConf} {\bf C0706044} (2007) 25},
  [\href{http://arxiv.org/abs/0710.5878}{{\tt 0710.5878}}].

\bibitem{Cvetic:2008bn}
G.~Cvetic and C.~Valenzuela, \emph{{Analytic QCD: A Short review}},
  {\emph{Braz. J. Phys.} {\bf 38} (2008) 371--380},
  [\href{http://arxiv.org/abs/0804.0872}{{\tt 0804.0872}}].

\bibitem{Peris:1998nj}
S.~Peris, M.~Perrottet and E.~de~Rafael, \emph{{Matching long and short
  distances in large N(c) QCD}},
  \href{http://dx.doi.org/10.1088/1126-6708/1998/05/011}{\emph{JHEP} {\bf 05}
  (1998) 011}, [\href{http://arxiv.org/abs/hep-ph/9805442}{{\tt
  hep-ph/9805442}}].

\bibitem{Maiezza:2021mry}
A.~Maiezza and J.~C. Vasquez, \emph{{Resurgence of the QCD Adler function}},
  \href{http://dx.doi.org/10.1016/j.physletb.2021.136338}{\emph{Phys. Lett. B}
  {\bf 817} (2021) 136338}, [\href{http://arxiv.org/abs/2104.03095}{{\tt
  2104.03095}}].

\bibitem{Ecalle1993}
J.~\'Ecalle, \emph{Six lectures on transseries, analysable functions and the
  constructive proof of dulac's conjecture}, .

\bibitem{Argyres:2012ka}
P.~C. Argyres and M.~Unsal, \emph{{The semi-classical expansion and resurgence
  in gauge theories: new perturbative, instanton, bion, and renormalon
  effects}}, \href{http://dx.doi.org/10.1007/JHEP08(2012)063}{\emph{JHEP} {\bf
  08} (2012) 063}, [\href{http://arxiv.org/abs/1206.1890}{{\tt 1206.1890}}].

\bibitem{Dunne:2012ae}
G.~V. Dunne and M.~Unsal, \emph{{Resurgence and Trans-series in Quantum Field
  Theory: The CP(N-1) Model}},
  \href{http://dx.doi.org/10.1007/JHEP11(2012)170}{\emph{JHEP} {\bf 11} (2012)
  170}, [\href{http://arxiv.org/abs/1210.2423}{{\tt 1210.2423}}].

\bibitem{Dorigoni:2014hea}
D.~Dorigoni, \emph{{An Introduction to Resurgence, Trans-Series and Alien
  Calculus}}, \href{http://dx.doi.org/10.1016/j.aop.2019.167914}{\emph{Annals
  Phys.} {\bf 409} (2019) 167914}, [\href{http://arxiv.org/abs/1411.3585}{{\tt
  1411.3585}}].

\bibitem{Aniceto:2018bis}
I.~Aniceto, G.~Basar and R.~Schiappa, \emph{{A Primer on Resurgent Transseries
  and Their Asymptotics}},
  \href{http://dx.doi.org/10.1016/j.physrep.2019.02.003}{\emph{Phys. Rept.}
  {\bf 809} (2019) 1--135}, [\href{http://arxiv.org/abs/1802.10441}{{\tt
  1802.10441}}].

\bibitem{Clavier:2019sph}
P.~J. Clavier, \emph{{Borel-Ecalle resummation of a two-point function}},
  \href{http://arxiv.org/abs/1912.03237}{{\tt 1912.03237}}.

\bibitem{Borinsky:2020vae}
M.~Borinsky and G.~V. Dunne, \emph{{Non-Perturbative Completion of
  Hopf-Algebraic Dyson-Schwinger Equations}},
  \href{http://dx.doi.org/10.1016/j.nuclphysb.2020.115096}{\emph{Nucl. Phys. B}
  {\bf 957} (2020) 115096}, [\href{http://arxiv.org/abs/2005.04265}{{\tt
  2005.04265}}].

\bibitem{Fujimori:2021oqg}
T.~Fujimori, M.~Honda, S.~Kamata, T.~Misumi, N.~Sakai and T.~Yoda,
  \emph{{Quantum phase transition and Resurgence: Lessons from 3d
  $\mathcal{N}=4$ SQED}},  \href{http://arxiv.org/abs/2103.13654}{{\tt
  2103.13654}}.

\bibitem{Costin:2019xql}
O.~Costin and G.~V. Dunne, \emph{{Resurgent extrapolation: rebuilding a
  function from asymptotic data. Painlev\'e I}},
  \href{http://dx.doi.org/10.1088/1751-8121/ab477b}{\emph{J. Phys. A} {\bf 52}
  (2019) 445205}, [\href{http://arxiv.org/abs/1904.11593}{{\tt 1904.11593}}].

\bibitem{Costin:2020hwg}
O.~Costin and G.~V. Dunne, \emph{{Physical Resurgent Extrapolation}},
  \href{http://dx.doi.org/10.1016/j.physletb.2020.135627}{\emph{Phys. Lett. B}
  {\bf 808} (2020) 135627}, [\href{http://arxiv.org/abs/2003.07451}{{\tt
  2003.07451}}].

\bibitem{Borinsky:2022knn}
M.~Borinsky and D.~Broadhurst, \emph{{Resonant resurgent asymptotics from
  quantum field theory}},  \href{http://arxiv.org/abs/2202.01513}{{\tt
  2202.01513}}.

\bibitem{Costin1995}
O.~Costin\href{http://dx.doi.org/10.1155/s1073792895000286}{\emph{International
  Mathematics Research Notices} {\bf 1995} (1995) 377}.

\bibitem{CostinBook}
O.~Costin, \emph{{Asymptotics and Borel Summability. Monographs and Surveys in
  Pure and Applied Mathematics. Chapman and Hall/CRC (2008)}}, .

\bibitem{Maiezza:2019dht}
A.~Maiezza and J.~C. Vasquez, \emph{{Non-local Lagrangians from Renormalons and
  Analyzable Functions}},
  \href{http://dx.doi.org/10.1016/j.aop.2019.04.015}{\emph{Annals Phys.} {\bf
  407} (2019) 78--91}, [\href{http://arxiv.org/abs/1902.05847}{{\tt
  1902.05847}}].

\bibitem{Bersini:2019axn}
J.~Bersini, A.~Maiezza and J.~C. Vasquez, \emph{{Resurgence of the
  Renormalization Group Equation}},
  \href{http://dx.doi.org/10.1016/j.aop.2020.168126}{\emph{Annals Phys.} {\bf
  415} (2020) 168126}, [\href{http://arxiv.org/abs/1910.14507}{{\tt
  1910.14507}}].

\bibitem{Landau}
L.~D. Landau, \emph{{\it Niels Bohr and the Development of Physics}},
  {\emph{{Pergamon Press. London}} (1955) }.

\bibitem{Cornwall:1981zr}
J.~M. Cornwall, \emph{{Dynamical Mass Generation in Continuum QCD}},
  \href{http://dx.doi.org/10.1103/PhysRevD.26.1453}{\emph{Phys. Rev. D} {\bf
  26} (1982) 1453}.

\bibitem{PhysRevD.44.1285}
J.~Papavassiliou and J.~M. Cornwall, \emph{Coupled fermion gap and vertex
  equations for chiral-symmetry breakdown in qcd},
  \href{http://dx.doi.org/10.1103/PhysRevD.44.1285}{\emph{Phys. Rev. D} {\bf
  44} (Aug, 1991) 1285--1297}.

\bibitem{Deur:2016tte}
A.~Deur, S.~J. Brodsky and G.~F. de~Teramond, \emph{{The QCD Running
  Coupling}}, \href{http://dx.doi.org/10.1016/j.ppnp.2016.04.003}{\emph{Nucl.
  Phys.} {\bf 90} (2016) 1}, [\href{http://arxiv.org/abs/1604.08082}{{\tt
  1604.08082}}].

\bibitem{Miller:2007kk}
J.~P. Miller, E.~de~Rafael and B.~L. Roberts, \emph{{Muon (g-2): Experiment and
  theory}}, \href{http://dx.doi.org/10.1088/0034-4885/70/5/R03}{\emph{Rept.
  Prog. Phys.} {\bf 70} (2007) 795},
  [\href{http://arxiv.org/abs/hep-ph/0703049}{{\tt hep-ph/0703049}}].

\bibitem{Miller:2012opa}
J.~P. Miller, E.~de~Rafael, B.~L. Roberts and D.~St\"ockinger, \emph{{Muon
  (g-2): Experiment and Theory}},
  \href{http://dx.doi.org/10.1146/annurev-nucl-031312-120340}{\emph{Ann. Rev.
  Nucl. Part. Sci.} {\bf 62} (2012) 237--264}.

\bibitem{Keshavarzi:2020bfy}
A.~Keshavarzi, W.~J. Marciano, M.~Passera and A.~Sirlin, \emph{{Muon $g-2$ and
  $\Delta \alpha$ connection}},
  \href{http://dx.doi.org/10.1103/PhysRevD.102.033002}{\emph{Phys. Rev. D} {\bf
  102} (2020) 033002}, [\href{http://arxiv.org/abs/2006.12666}{{\tt
  2006.12666}}].

\bibitem{PhysRevLett.126.141801}
{\scshape Muon $g\ensuremath{-}2$ Collaboration} collaboration, B.~Abi,
  T.~Albahri, S.~Al-Kilani, D.~Allspach, L.~P. Alonzi, A.~Anastasi et~al.,
  \emph{Measurement of the positive muon anomalous magnetic moment to 0.46
  ppm}, \href{http://dx.doi.org/10.1103/PhysRevLett.126.141801}{\emph{Phys.
  Rev. Lett.} {\bf 126} (Apr, 2021) 141801}.

\bibitem{Borsanyi:2020mff}
S.~Borsanyi et~al., \emph{{Leading hadronic contribution to the muon magnetic
  moment from lattice QCD}},
  \href{http://dx.doi.org/10.1038/s41586-021-03418-1}{\emph{Nature} {\bf 593}
  (2021) 51--55}, [\href{http://arxiv.org/abs/2002.12347}{{\tt 2002.12347}}].

\bibitem{Gorishnii:1990vf}
S.~G. Gorishnii, A.~L. Kataev and S.~A. Larin, \emph{{The
  $O(\alpha^{3}_{s})$-corrections to $\sigma_{tot}(e^{+}e^{-}\rightarrow
  hadrons)$ and $\Gamma(\tau^{-} \rightarrow \nu_{\tau} + hadrons)$ in QCD}},
  \href{http://dx.doi.org/10.1016/0370-2693(91)90149-K}{\emph{Phys. Lett. B}
  {\bf 259} (1991) 144--150}.

\bibitem{Surguladze:1990tg}
L.~R. Surguladze and M.~A. Samuel, \emph{{Total hadronic cross-section in e+ e-
  annihilation at the four loop level of perturbative QCD}},
  \href{http://dx.doi.org/10.1103/PhysRevLett.66.560}{\emph{Phys. Rev. Lett.}
  {\bf 66} (1991) 560--563}.

\bibitem{Kataev:1995vh}
A.~L. Kataev and V.~V. Starshenko, \emph{{Estimates of the higher order QCD
  corrections to R(s), R(tau) and deep inelastic scattering sum rules}},
  \href{http://dx.doi.org/10.1142/S0217732395000272}{\emph{Mod. Phys. Lett. A}
  {\bf 10} (1995) 235--250}, [\href{http://arxiv.org/abs/hep-ph/9502348}{{\tt
  hep-ph/9502348}}].

\bibitem{Beneke:1994qe}
M.~Beneke and V.~M. Braun, \emph{{Naive nonAbelianization and resummation of
  fermion bubble chains}},
  \href{http://dx.doi.org/10.1016/0370-2693(95)00184-M}{\emph{Phys. Lett. B}
  {\bf 348} (1995) 513--520}, [\href{http://arxiv.org/abs/hep-ph/9411229}{{\tt
  hep-ph/9411229}}].

\bibitem{Lipatov:1976rb}
L.~N. Lipatov, \emph{{Multi-Regge Processes and the Pomeranchuk Singularity in
  Nonabelian Gauge Theories}},  in \emph{{Proceedings, XVIII International
  Conference on High-Energy Physics Volume 1: July 15-21, 1976 Tbilisi, USSR}},
  pp.~A5.26--28, 1976.

\bibitem{Maiezza:2020qib}
A.~Maiezza and J.~C. Vasquez, \emph{{On Haag\textquoteright{}s Theorem and
  Renormalization Ambiguities}},
  \href{http://dx.doi.org/10.1007/s10701-021-00484-3}{\emph{Found. Phys.} {\bf
  51} (2021) 80}, [\href{http://arxiv.org/abs/2011.08875}{{\tt 2011.08875}}].

\bibitem{Neubert:1994vb}
M.~Neubert, \emph{{Scale setting in QCD and the momentum flow in Feynman
  diagrams}}, \href{http://dx.doi.org/10.1103/PhysRevD.51.5924}{\emph{Phys.
  Rev. D} {\bf 51} (1995) 5924--5941},
  [\href{http://arxiv.org/abs/hep-ph/9412265}{{\tt hep-ph/9412265}}].

\bibitem{Dokshitzer:1995qm}
Y.~L. Dokshitzer, G.~Marchesini and B.~R. Webber, \emph{{Dispersive approach to
  power behaved contributions in QCD hard processes}},
  \href{http://dx.doi.org/10.1016/0550-3213(96)00155-1}{\emph{Nucl. Phys. B}
  {\bf 469} (1996) 93--142}, [\href{http://arxiv.org/abs/hep-ph/9512336}{{\tt
  hep-ph/9512336}}].

\bibitem{Parisi:1978bj}
G.~Parisi, \emph{{Singularities of the Borel Transform in Renormalizable
  Theories}}, \href{http://dx.doi.org/10.1016/0370-2693(78)90101-6}{\emph{Phys.
  Lett. B} {\bf 76} (1978) 65--66}.

\bibitem{Dokshitzer:1995af}
Y.~L. Dokshitzer and N.~G. Uraltsev, \emph{{Are IR renormalons a good probe for
  the strong interaction domain?}},
  \href{http://dx.doi.org/10.1016/0370-2693(96)00476-5}{\emph{Phys. Lett. B}
  {\bf 380} (1996) 141--150}, [\href{http://arxiv.org/abs/hep-ph/9512407}{{\tt
  hep-ph/9512407}}].

\bibitem{Peris:1996in}
S.~Peris and E.~de~Rafael, \emph{{On renormalons and Landau poles in gauge
  field theories}},
  \href{http://dx.doi.org/10.1016/0370-2693(96)01053-2}{\emph{Phys. Lett. B}
  {\bf 387} (1996) 603--608}, [\href{http://arxiv.org/abs/hep-ph/9603359}{{\tt
  hep-ph/9603359}}].

\bibitem{Maiezza:2020nbe}
A.~Maiezza and J.~C. Vasquez, \emph{{Non-Wilsonian ultraviolet completion via
  transseries}}, \href{http://dx.doi.org/10.1142/S0217751X21500160}{\emph{Int.
  J. Mod. Phys. A} {\bf 36} (2021) 2150016},
  [\href{http://arxiv.org/abs/2007.01270}{{\tt 2007.01270}}].

\bibitem{Klaczynski:2013fca}
L.~Klaczynski and D.~Kreimer, \emph{{Avoidance of a Landau Pole by Flat
  Contributions in QED}},
  \href{http://dx.doi.org/10.1016/j.aop.2014.02.019}{\emph{Annals Phys.} {\bf
  344} (2014) 213--231}, [\href{http://arxiv.org/abs/1309.5061}{{\tt
  1309.5061}}].

\bibitem{Cvetic:2006mk}
G.~Cvetic and C.~Valenzuela, \emph{{An Approach for evaluation of observables
  in analytic versions of QCD}},
  \href{http://dx.doi.org/10.1088/0954-3899/32/6/L01}{\emph{J. Phys. G} {\bf
  32} (2006) L27}, [\href{http://arxiv.org/abs/hep-ph/0601050}{{\tt
  hep-ph/0601050}}].

\bibitem{Cvetic:2006gc}
G.~Cvetic and C.~Valenzuela, \emph{{Various versions of analytic QCD and
  skeleton-motivated evaluation of observables}},
  \href{http://dx.doi.org/10.1103/PhysRevD.74.114030}{\emph{Phys. Rev. D} {\bf
  74} (2006) 114030}, [\href{http://arxiv.org/abs/hep-ph/0608256}{{\tt
  hep-ph/0608256}}].

\bibitem{Cvetic:2013gta}
G.~Cvetic, \emph{{Techniques of evaluation of QCD low-energy physical
  quantities with running coupling with infrared fixed point}},
  \href{http://dx.doi.org/10.1103/PhysRevD.89.036003}{\emph{Phys. Rev. D} {\bf
  89} (2014) 036003}, [\href{http://arxiv.org/abs/1309.1696}{{\tt 1309.1696}}].

\bibitem{Davier:2005xq}
M.~Davier, A.~Hocker and Z.~Zhang, \emph{{The Physics of Hadronic Tau Decays}},
  \href{http://dx.doi.org/10.1103/RevModPhys.78.1043}{\emph{Rev. Mod. Phys.}
  {\bf 78} (2006) 1043--1109}, [\href{http://arxiv.org/abs/hep-ph/0507078}{{\tt
  hep-ph/0507078}}].

\bibitem{Czarnecki:2000id}
A.~Czarnecki and W.~J. Marciano, \emph{{The Muon anomalous magnetic moment:
  Standard model theory and beyond}},  in \emph{{5th International Symposium on
  Radiative Corrections: Applications of Quantum Field Theory to
  Phenomenology}}, 9, 2000.
\newblock \href{http://arxiv.org/abs/hep-ph/0010194}{{\tt hep-ph/0010194}}.

\bibitem{Czarnecki:2001pv}
A.~Czarnecki and W.~J. Marciano, \emph{{The Muon anomalous magnetic moment: A
  Harbinger for 'new physics'}},
  \href{http://dx.doi.org/10.1103/PhysRevD.64.013014}{\emph{Phys. Rev. D} {\bf
  64} (2001) 013014}, [\href{http://arxiv.org/abs/hep-ph/0102122}{{\tt
  hep-ph/0102122}}].

\bibitem{Knecht:2003kc}
M.~Knecht, \emph{{The Anomalous magnetic moment of the muon: A Theoretical
  introduction}},
  \href{http://dx.doi.org/10.1007/978-3-540-44457-2_2}{\emph{Lect. Notes Phys.}
  {\bf 629} (2004) 37--84}, [\href{http://arxiv.org/abs/hep-ph/0307239}{{\tt
  hep-ph/0307239}}].

\bibitem{Davier:2004gb}
M.~Davier and W.~J. Marciano, \emph{{The theoretical prediction for the muon
  anomalous magnetic moment}},
  \href{http://dx.doi.org/10.1146/annurev.nucl.54.070103.181204}{\emph{Ann.
  Rev. Nucl. Part. Sci.} {\bf 54} (2004) 115--140}.

\bibitem{Jegerlehner:2009ry}
F.~Jegerlehner and A.~Nyffeler, \emph{{The Muon g-2}},
  \href{http://dx.doi.org/10.1016/j.physrep.2009.04.003}{\emph{Phys. Rept.}
  {\bf 477} (2009) 1--110}, [\href{http://arxiv.org/abs/0902.3360}{{\tt
  0902.3360}}].

\bibitem{Aoyama:2020ynm}
T.~Aoyama et~al., \emph{{The anomalous magnetic moment of the muon in the
  Standard Model}},
  \href{http://dx.doi.org/10.1016/j.physrep.2020.07.006}{\emph{Phys. Rept.}
  {\bf 887} (2020) 1--166}, [\href{http://arxiv.org/abs/2006.04822}{{\tt
  2006.04822}}].

\bibitem{Cvetic:2020naz}
G.~Cveti\v{c} and R.~K\"ogerler, \emph{{Lattice-motivated QCD coupling and
  hadronic contribution to muon $g - 2$}},
  \href{http://dx.doi.org/10.1088/1361-6471/abd259}{\emph{J. Phys. G} {\bf 48}
  (2021) 055008}, [\href{http://arxiv.org/abs/2009.13742}{{\tt 2009.13742}}].

\bibitem{Davier:2010nc}
M.~Davier, A.~Hoecker, B.~Malaescu and Z.~Zhang, \emph{{Reevaluation of the
  Hadronic Contributions to the Muon g-2 and to alpha(MZ)}},
  \href{http://dx.doi.org/10.1140/epjc/s10052-012-1874-8}{\emph{Eur. Phys. J.
  C} {\bf 71} (2011) 1515}, [\href{http://arxiv.org/abs/1010.4180}{{\tt
  1010.4180}}].

\bibitem{Davier:2017zfy}
M.~Davier, A.~Hoecker, B.~Malaescu and Z.~Zhang, \emph{{Reevaluation of the
  hadronic vacuum polarisation contributions to the Standard Model predictions
  of the muon $g-2$ and ${\alpha (m_Z^2)}$ using newest hadronic cross-section
  data}}, \href{http://dx.doi.org/10.1140/epjc/s10052-017-5161-6}{\emph{Eur.
  Phys. J. C} {\bf 77} (2017) 827},
  [\href{http://arxiv.org/abs/1706.09436}{{\tt 1706.09436}}].

\bibitem{Davier:2019can}
M.~Davier, A.~Hoecker, B.~Malaescu and Z.~Zhang, \emph{{A new evaluation of the
  hadronic vacuum polarisation contributions to the muon anomalous magnetic
  moment and to $\mathbf{\boldsymbol\alpha(m_Z^2)}$}},
  \href{http://dx.doi.org/10.1140/epjc/s10052-020-7792-2}{\emph{Eur. Phys. J.
  C} {\bf 80} (2020) 241}, [\href{http://arxiv.org/abs/1908.00921}{{\tt
  1908.00921}}].

\bibitem{Lautrup:1971jf}
B.~e. Lautrup, A.~Peterman and E.~de~Rafael, \emph{{Recent developments in the
  comparison between theory and experiments in quantum electrodynamics}},
  \href{http://dx.doi.org/10.1016/0370-1573(72)90011-7}{\emph{Phys. Rept.} {\bf
  3} (1972) 193--259}.

\bibitem{PhysRevLett.125.091801}
A.~Crivellin, M.~Hoferichter, C.~A. Manzari and M.~Montull, \emph{Hadronic
  vacuum polarization: $(g\ensuremath{-}2{)}_{\ensuremath{\mu}}$ versus global
  electroweak fits},
  \href{http://dx.doi.org/10.1103/PhysRevLett.125.091801}{\emph{Phys. Rev.
  Lett.} {\bf 125} (Aug, 2020) 091801}.

\bibitem{Malaescu:2020zuc}
B.~Malaescu and M.~Schott, \emph{{Impact of correlations between $a_{\mu }$ and
  $\alpha _\text {QED}$ on the EW fit}},
  \href{http://dx.doi.org/10.1140/epjc/s10052-021-08848-9}{\emph{Eur. Phys. J.
  C} {\bf 81} (2021) 46}, [\href{http://arxiv.org/abs/2008.08107}{{\tt
  2008.08107}}].

\bibitem{Colangelo:2020lcg}
G.~Colangelo, M.~Hoferichter and P.~Stoffer, \emph{{Constraints on the two-pion
  contribution to hadronic vacuum polarization}},
  \href{http://dx.doi.org/10.1016/j.physletb.2021.136073}{\emph{Phys. Lett. B}
  {\bf 814} (2021) 136073}, [\href{http://arxiv.org/abs/2010.07943}{{\tt
  2010.07943}}].

\bibitem{Banerjee:2020tdt}
P.~Banerjee et~al., \emph{{Theory for muon-electron scattering @ 10 ppm: A
  report of the MUonE theory initiative}},
  \href{http://dx.doi.org/10.1140/epjc/s10052-020-8138-9}{\emph{Eur. Phys. J.
  C} {\bf 80} (2020) 591}, [\href{http://arxiv.org/abs/2004.13663}{{\tt
  2004.13663}}].

\bibitem{Webber:1994cp}
B.~R. Webber, \emph{{Estimation of power corrections to hadronic event
  shapes}}, \href{http://dx.doi.org/10.1016/0370-2693(94)91147-9}{\emph{Phys.
  Lett. B} {\bf 339} (1994) 148--150},
  [\href{http://arxiv.org/abs/hep-ph/9408222}{{\tt hep-ph/9408222}}].

\bibitem{Manohar:1994kq}
A.~V. Manohar and M.~B. Wise, \emph{{Power suppressed corrections to hadronic
  event shapes}},
  \href{http://dx.doi.org/10.1016/0370-2693(94)01504-6}{\emph{Phys. Lett. B}
  {\bf 344} (1995) 407--412}, [\href{http://arxiv.org/abs/hep-ph/9406392}{{\tt
  hep-ph/9406392}}].

\bibitem{Korchemsky:1994is}
G.~P. Korchemsky and G.~F. Sterman, \emph{{Nonperturbative corrections in
  resummed cross-sections}},
  \href{http://dx.doi.org/10.1016/0550-3213(94)00006-Z}{\emph{Nucl. Phys. B}
  {\bf 437} (1995) 415--432}, [\href{http://arxiv.org/abs/hep-ph/9411211}{{\tt
  hep-ph/9411211}}].

\bibitem{Dokshitzer:1995zt}
Y.~L. Dokshitzer and B.~R. Webber, \emph{{Calculation of power corrections to
  hadronic event shapes}},
  \href{http://dx.doi.org/10.1016/0370-2693(95)00548-Y}{\emph{Phys. Lett. B}
  {\bf 352} (1995) 451--455}, [\href{http://arxiv.org/abs/hep-ph/9504219}{{\tt
  hep-ph/9504219}}].

\bibitem{Akhoury:1995sp}
R.~Akhoury and V.~I. Zakharov, \emph{{On the universality of the leading, 1/Q
  power corrections in QCD}},
  \href{http://dx.doi.org/10.1016/0370-2693(95)00866-J}{\emph{Phys. Lett. B}
  {\bf 357} (1995) 646--652}, [\href{http://arxiv.org/abs/hep-ph/9504248}{{\tt
  hep-ph/9504248}}].

\bibitem{Nason:1995np}
P.~Nason and M.~H. Seymour, \emph{{Infrared renormalons and power suppressed
  effects in e+ e- jet events}},
  \href{http://dx.doi.org/10.1016/0550-3213(95)00461-Z}{\emph{Nucl. Phys. B}
  {\bf 454} (1995) 291--312}, [\href{http://arxiv.org/abs/hep-ph/9506317}{{\tt
  hep-ph/9506317}}].

\bibitem{Beneke:1994sw}
M.~Beneke and V.~M. Braun, \emph{{Heavy quark effective theory beyond
  perturbation theory: Renormalons, the pole mass and the residual mass term}},
  \href{http://dx.doi.org/10.1016/0550-3213(94)90314-X}{\emph{Nucl. Phys. B}
  {\bf 426} (1994) 301--343}, [\href{http://arxiv.org/abs/hep-ph/9402364}{{\tt
  hep-ph/9402364}}].

\bibitem{Bigi:1994em}
I.~I.~Y. Bigi, M.~A. Shifman, N.~G. Uraltsev and A.~I. Vainshtein, \emph{{The
  Pole mass of the heavy quark. Perturbation theory and beyond}},
  \href{http://dx.doi.org/10.1103/PhysRevD.50.2234}{\emph{Phys. Rev. D} {\bf
  50} (1994) 2234--2246}, [\href{http://arxiv.org/abs/hep-ph/9402360}{{\tt
  hep-ph/9402360}}].

\bibitem{Aglietti:1995tg}
U.~Aglietti and Z.~Ligeti, \emph{{Renormalons and confinement}},
  \href{http://dx.doi.org/10.1016/0370-2693(95)01234-2}{\emph{Phys. Lett. B}
  {\bf 364} (1995) 75}, [\href{http://arxiv.org/abs/hep-ph/9503209}{{\tt
  hep-ph/9503209}}].

\bibitem{costin1998}
O.~Costin, \emph{On borel summation and stokes phenomena for rank- $1$
  nonlinear systems of ordinary differential equations},
  \href{http://dx.doi.org/10.1215/S0012-7094-98-09311-5}{\emph{Duke Math. J.}
  {\bf 93} (06, 1998) 289--344}.

\bibitem{Broadhurst:1992si}
D.~J. Broadhurst, \emph{{Large N expansion of QED: Asymptotic photon propagator
  and contributions to the muon anomaly, for any number of loops}},
  \href{http://dx.doi.org/10.1007/BF01560355}{\emph{Z. Phys. C} {\bf 58} (1993)
  339--346}.

\end{thebibliography}\endgroup

\end{document}